\documentclass{aa}
\newcommand{\kms}{km~s$^{-1}$}
\usepackage{graphicx}
\begin{document}
\title{Abundances for metal-poor stars with accurate parallaxes
\thanks{Based in part on data collected at the European Southern
Observatory, Chile, at the MacDonald Observatory, Texas, USA, and at the
Telescopio Nazionale Galileo, Canary Island, INAF, Italy-Spain}}
\subtitle{II. $\alpha-$elements in the halo}

\author{R.G. Gratton\inst{1}, 
E. Carretta\inst{1}, 
S. Desidera\inst{1},
S. Lucatello\inst{1,2},
P. Mazzei\inst{1},
and M. Barbieri\inst{3}}

\offprints{R.G. Gratton}

\institute{INAF-Osservatorio Astronomico di Padova, Vicolo dell'Osservatorio
5, 35122 Padova, Italy\\
\and
Dipartimento di Astronomia, Universit\`a di Padova, Italy
\and
CISAS, Universit\a di Padova, Italy}

\date{Received March, 17 2003; accepted May, 13 2003}

\abstract{ Abundances for $\alpha-$elements and Fe in about 150 field
subdwarfs and early subgiants with accurate parallaxes and kinematic data are
used to discuss the run of abundance ratios in metal-poor stars in the solar
neighborhood. Based on kinematics, we separated stars into two populations:
the first one has a positive velocity of rotation around the galactic center,
and it is likely to be related to the dissipational collapse of the galaxy;
the second one has either negligible or negative rotational velocity, and it
is likely related to an accretion component. The two populations show a large
overlap in metallicity. However, they show distinct chemical properties. For
the first population we found that there are close correlations (with small
scatters around) of the rotational velocity with metallicity and with the
Fe/$\alpha$\ abundance ratio: this might be a signature of a not very fast
collapse of the progenitor clouds, with enough time for a significant
contribution by SNe Ia, although this result needs to be confirmed by a
3-D/non-LTE study. On the other side, the second population exhibits a
larger scatter in both the above mentioned relations, and on average, a larger
Fe/$\alpha$\ ratio at a given metallicity, suggesting a larger scatter in
ages. We argue that the lack of stars with moderate rotational velocities and
high Fe/$\alpha$\ abundance ratios is due to the short merging time for
protogalactic clouds with prograde motion, while the presence of a group of
counter-rotating stars with this characteristics indicates a much longer
typical lifetimes for protogalactic fragments having such a motion. Finally,
we found that perigalactic distances correlate with the Fe/$\alpha$\ abundance
ratios better than the apogalactic distances.
     \keywords{ Stars: abundances --
                 Stars: evolution --
                 Stars: Population II --
            	 Galaxy: globular clusters: general
               }
   }

\authorrunning{Gratton R.G. et al.}
\titlerunning{Abundances in metal poor-stars}

   \maketitle
%
%________________________________________________________________

\section{Introduction}

Galaxy formation theory must account for the wide diversity of observed
structures, from dwarf irregulars to grand design spirals, from low brightness
dwarfs to luminous ellipticals. This involves at least two critical
ingredients: the formation of dark halos and global formation of stars. The
hierarchical clustering is currently the most successful paradigm of
structures formation. In the past two decades numerical simulations have
become an important tool for exploring the detailed predictions of these
models (Weinberg et al. 2002). However, extending these studies to understand
the formation of luminous components of galaxies is difficult since our
knowledge of the star formation  process and its interaction (feedback) with
the surrounding interstellar medium is still rather limited (Kay et al. 2002).
Local Group galaxies, and in particular our own Galaxy, represent a unique
opportunity to tune these models. Recently, given the improved  capabilities
of new generation computers, which allow us to carry out very high particle
resolution runs, the first numerical  simulations of disk dominated  galaxy
formation in a fully consistent cosmological framework have been performed
(Samland \& Gerhard, 2003; Abadi et al. 2002; Governato et al. 2002). There
are, however, many unresolved problems affecting both  dark matter and
galaxies  in such a framework (Steinmez, 2001; Silk \& Bouwens, 2001).
Detailed studies of the Galaxy can help our understanding of the complex
processes involved in baryons dissipation (see e.g. the review by Freeman \&
Bland-Hawthorn, 2002). Eggen et al. (1962) were the first to show that it is
possible to study Galaxy evolution using stellar abundances and stellar
dynamics. They derive that metal--poor stars are in the halo that was created
during the rapid collapse of a relatively uniform, isolated protogalactic
cloud shortly after it decoupled from the universal expansion. In this
scenario, gradients of metallicity and age with respect to kinematics, as
represented by rotational velocity or eccentricity, should be present (see
e.g. Larson 1974, 1976), and the morphology of a galaxy should be related to
the activity of star formation in early epochs compared to the collapse time.
Angular momentum conservation causes the evolution of the residual gas from an
original roughly spherical distribution toward a disk. This picture was
challenged by  Searle \& Zinn (1978). They found that the galactic globular
clusters have a wide range of metal abundances independent of their radius
from the galactic center. Their findings pointed out that individual protogalactic
clouds might have had a significant independent chemical and dynamical
evolution before or during their merging into the main galactic body. Relics
of this independent evolution may be traced by careful and extensive studies
of
 chemical and dynamical properties of the earlier galactic stellar
generations. The ideas of galaxy formation via hierarchical
aggregation/merging of smaller units from the early Universe well agrees with
the Searle and Zinn scenario of the galactic halo formation. However within
this general framework, the role of the Milky Way is still to be adequately
clarified. The Milky Way is a spiral galaxy located in a region of relatively
low overdensity, at the edges of the Local Supercluster, with a quite
significant  bulge, a rather extended thin disk, a thick disk (Gilmore et al.
1989) and a halo. The presence of a dominant stellar (and gaseous) disk
suggests that dissipational collapse, the main mechanism in the formation of
spiral systems, had played a very important role. Moreover there are several
observational facts that do not fit well in a monolithic dissipational
collapse scheme. First, stars in the disk do not show the expected run of
increasing metallicity with increasing age; also there is a lack of metal-poor
stars (the so-called G-dwarf problem). Both these unexpected features can be
understood by assuming that there is a continuous infall of metal-poor
material onto the disk. The origin of this material is not  clear (although a number of
interstellar clouds at high latitudes have been observed since a long time,
starting with IRAS satellite). Second, the real differences between the
different components of our Galaxy are still to be properly understood.
Gratton et al. (1996, 2000) and Fuhrmann (1998) have shown that the ratio between
$\alpha-$elements (almost exclusively produced by short living core collapse
Supernovae- SNe hereinafter) and Fe (a rather large fraction of which is
produced on much longer time scale by thermonuclear SNe) indicates a clear
distinction between the epoch of formation of the thin and the thick disk,
with a phase of low star formation in between. This has led several authors to
construct two-infall models, that reproduce quite well these observed features
(see e.g. Chiappini et al. 1997). However, the reason for this hiatus in the
star formation is still unclear. It may be due to either a  merger event in
the early history of our Galaxy,
 perhaps related to the formation of the
bulge of our Galaxy, or the result of a galactic wind caused by a large number
of SNe explosions in the later phases of the thick disk formation (or perhaps,
to both these mechanisms).

On the other side, the thick disk seems to have properties that distinguish it
from the halo too: in particular, it seems much more chemically homogeneous
(see Nissen \& Schuster 1997). It should be noted that the halo itself is
often considered to be composed of two distinct populations, one with some
rotation, and the other either not-rotating at all, or even counter-rotating
(Norris 1994). The second of these populations probably traces accretion
events that occurred later in the galactic history; we are even actually
observing one of these merging fragment, the Sagittarius dwarf Galaxy (Ibata
et al. 1995). The relation between the rotating component of the halo and the
thick disk is unclear, it may well actually be a single population with a
continuum of properties.

In this paper we present a discussion of the abundances of $\alpha-$elements
and Fe for a group of 150 metal-poor stars with accurate Hipparcos parallaxes,
whose analysis was presented in a previous paper (Gratton et al. 2003,
hereinafter Paper I). High precision abundances and kinematics were obtained
for these stars: this allows us to discuss some properties of the the halo. In
particular, we do not find any discontinuity between the properties of the
rotating component of the halo and the thick disk. They agree fairly well with
predictions for a dissipational collapse of a large fraction of the primordial
galaxy. On the other side, the not rotating component of the halo has clearly
distinct chemical properties, in agreement with scenarios where it results
from  accretion of protogalactic fragments with independent star formation
history and chemical evolution.

The paper is organized as follows: in Sect. 2 we recall the most relevant
points of Paper I (sample selection and derivation of kinematical data and
chemical composition); in Sect. 3 we present the abundances, while in
Sect. 4 we correlate these results with kinematical data and recall the most
important biases present in our sample. Finally, in Sect. 5 we discuss our
results in the framework of the early evolution of our Galaxy and present our
conclusions.

\section{Sample and analysis}

Details about the analysis are described in Paper I. Briefly, we considered a
sample of 146 mostly metal-poor stars with accurate parallaxes $\pi$\ ($\Delta
\pi/\pi<0.2$) from the Hipparcos catalogue (Perryman et al. 1997); accurate
space velocities could be derived for these stars, combining Hipparcos proper
motions and distances with radial velocities from the literature. To keep the
analysis as homogeneous as possible, we did not consider stars evolved off the
base of the subgiant branch. High precision equivalent widths (with errors
smaller than 3~m\AA) from high $S/N$, high resolution spectra were available
for these stars, either from our own observations (using UVES at VLT, SARG at
TNG, or the 2.7m McDonald telescope), or from the literature (Nissen \&
Schuster 1997; Fulbright 2000; Prochaska et al. 2000).

The stars were subdivided into three groups according to kinematic criteria
based on the results of appropriate Galactic orbit calculations. The groups
considered are:
\begin{enumerate}
\item A rotating inner population, with a galactic rotation velocity
larger than 40\,\kms\, and $R_{\rm max}<15$~kpc, where $R_{\rm max}$ is
the apogalactic distance. This population includes part of what usually is
called the Halo as well as the Thick Disk, and can be substantially identified
with the dissipative collapse population of Norris (1994), and with the
population first identified by Eggen et al. (1962). Hereinafter, we will refer
to this population as Dissipative Collapse Component. The reason for our
choice is that we are not be able to find in our data any clear discontinuity
between the properties of the rotating part of the Halo and those of the Thick
Disk.
\item A second population composed of non-rotating and counter-rotating stars.
This population includes the remaining part of what is usually called the
Halo, and can be substantially identified with the population due to accretion
processes first proposed by Searle \& Zinn (1978). As we will discuss in the
next sections, this population (as a group) has a chemical composition clearly
distinct from that of the first one, likely due to a different origin. We will
call these Accretion Component Stars.
\item Finally the Thin Disk, which also has clearly distinct chemical
composition from the first population, as first showed by Gratton et al.
(1996, 2000), and Fuhrmann (1998). The discontinuity in chemical composition
between the Dissipative Component and the Thin Disk is likely due to a phase
of low star formation that occurred during the early evolution of our Galaxy
(see also Chiappini et al. 1997). For such a component:
\begin{equation}
\sqrt{Z_{max}^2+4~e^2}<0.35,
\end{equation}
where $Z_{max}$ is the maximum height above the galactic plane (in kpc) and
{\it e} the eccentricity of the galactic orbit of the star. With respect
to the definitions of Paper I, we dropped the lower limit to [Fe/H] for the
thin disk population. In this way we have a pure kinematical definition of the
various populations. Two stars were attributed to the thin disk rather than to
the thick disk by dropping this criterion: HD 116316 ([Fe/H]=-0.71,
[$\alpha$/Fe]=0.19) and HD134169 (([Fe/H]=-0.84, [$\alpha$/Fe]=0.28). While
the impact of this change on the discussion is modest, we emphasize that
stellar populations are defined on the basis of statistical criteria, and
membership of individual stars may be questioned.
\end{enumerate}

A part from model uncertainties (which mainly affect absolute values, but only
marginally the relative ranking), the largest source of errors in our
kinematic is the error in the transverse motion related to the parallax. The
mean quadratic error in the parallaxes of the stars of our sample is of
$\Delta\pi/\pi=0.095$. Since the average transverse motion is of 141~\kms,
the typical error in the transverse motion is of about 14~\kms. This is much
larger than the typical error in radial velocities, that for the stars in our
sample is a few \kms. In order to estimate errors in the kinematical
parameters, we then set up the following Monte Carlo procedure. For each star 
we randomly extracted a series of values for the observational parameters 
(parallaxes, proper motions and radial velocities) obtained from Gaussian
distributions centered on the observed values and with dispersions equal to
the errors assumed for these quantities. Errors in the parallaxes are those
from the Hipparcos catalogue; for the remaining quantities we assumed typical
values of 1~mas/yr for the proper motions and 2~km/s for the radial velocities. 
With this new set of data, we run our orbit extraction parameters. We
repeated this procedure 100 times for each star, and adopted as errors in the
orbital parameters the standard deviation of the distribution of the values
obtained from each extraction. With this procedure we obtain that typical
observational errors on the kinematic parameters we consider in this
paper are $\pm 0.39$~kpc in the perigalactic distance $R_{\rm min}$, $\pm
0.028$~dex in the logarithm of the apogalactic distance $\log R_{\rm max}$, 
$\pm 0.046$\ in the eccentricity $e$, and $\pm 12$\,\kms\ in the rotational 
velocity $V_{\rm rot}$. Maximum errors (that is, errors for the star which
is most sensitive to possible uncertainties in the observational parameters)
are about twice as large. Such errors are small enough to have a
negligible impact on our discussion.

A quite standard abundance analysis was carried out using these data; however,
we carefully kept the star-to-star errors as small as possible.
Effective temperatures for the program stars were obtained using $B-V$\ and
$b-y$\ colors (generally assuming that interstellar reddening is negligible);
these were calibrated against effective temperature using the H$\alpha$\
profiles that were available for about a third of the stars in the sample. A
comparison with temperatures derived using the Infrared Flux Method by Alonso
et al. (1996) shows an excellent agreement, with only a few slightly
deviating point, generally due to stars in binary systems. The final adopted
temperatures should have typical errors of about 50\,K. Surface gravities were
obtained from the location of the stars in the color-magnitude diagrams, and
should have errors of about $\pm 0.1$\,dex. Finally, microturbulent velocities
were obtained minimizing any trend of derived abundances from expected
equivalent widths for Fe~I, and have errors of about $\pm 0.3$\,\kms.
Accurate, updated line parameters were considered: in particular, collisional
damping was considered using the recent treatment by Barklem et al. (2000),
that should be much superior to the classical method based on enhancement
factors over the classical van der Waals broadening given by the Uns\"old
formula. Moreover, oscillator strengths for Mg were carefully revised, and
non-LTE corrections were taken into account for O and Na lines.
On the other side, we do not find any evidence for significant departures
from LTE for the Fe lines used in our analysis.

Finally, two features must be mentioned. First, our analysis uses standard
plane-parallel atmosphere models by Kurucz (1994, with the overshooting option
switched off). We found a number of small but significant discrepancies in our
results that can be ascribed to limitations in these model atmospheres; such
limitations can perhaps be overcome using 3-d atmosphere models, with a much
more realistic treatment of convection, such as those recently used by Nissen
et al. (2002). Second, a quite significant fraction of the program stars are 
members of binary systems. Analysis for such stars is less accurate than for
single stars: in particular, significant errors can be present in the adopted
effective temperatures. As evidenced in Nissen et al. (2002) and in Paper I,
both these concerns are important in the analysis of some key elements, like 
Oxygen. However, the impact on the elements considered in this paper is much
smaller: for instance, we did not find any significant difference between
the [Fe/$\alpha$] ratios for bona fide single stars and for stars known to
be members of binary systems (at variance from what we found for the [O/Fe]
ratio). 

\begin{figure*} 
\includegraphics[width=13cm]{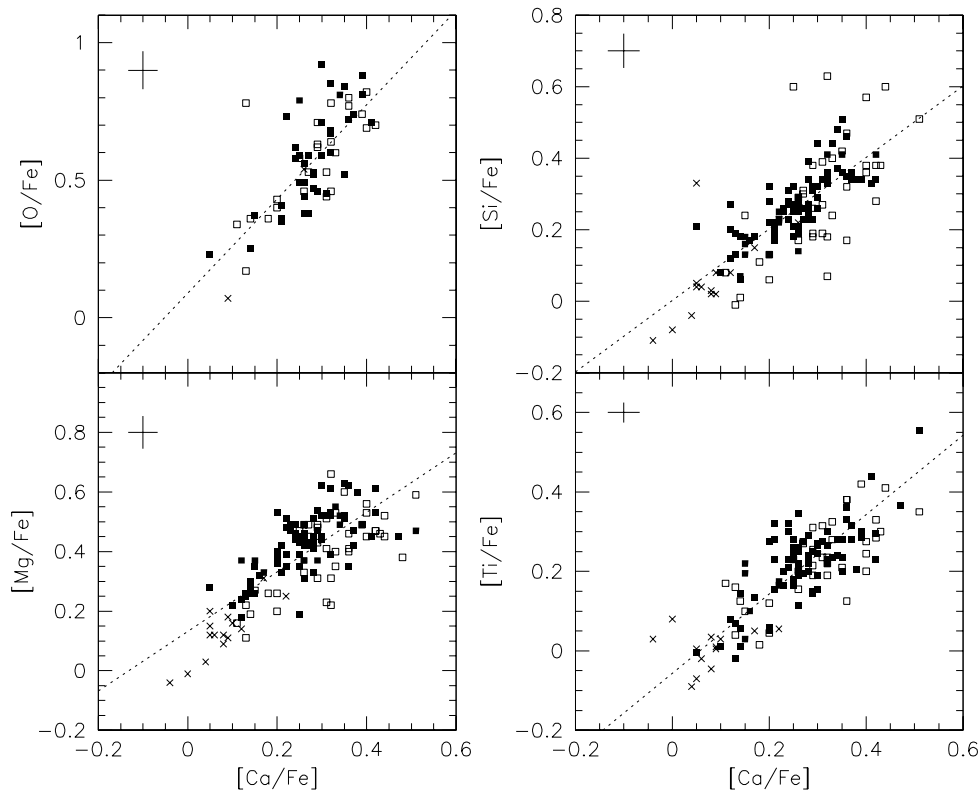}
\caption[]{ Comparison between the overabundances with respect to Fe for
the individual $\alpha-$elements with those obtained for Ca.
Filled squares are data for Dissipative Component stars;
open square those for the Accretion Component stars;
crosses are data for Thin Disk stars.
Error bars are shown in the upper left corner of each panel}
\label{f:fig00}
\end{figure*}

\begin{figure} 
\includegraphics[width=8.8cm]{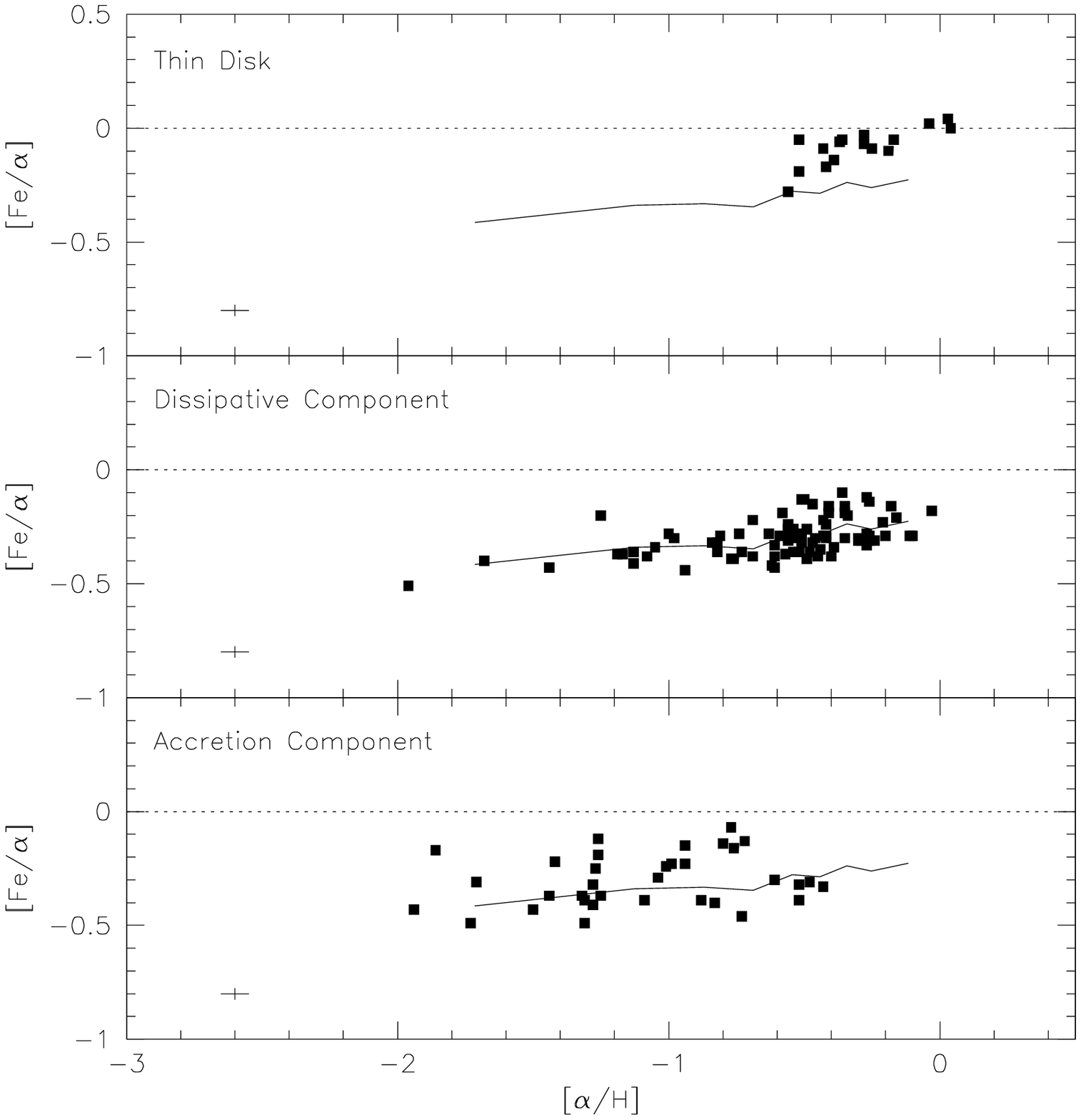}
\caption[]{ Run of the abundance ratio of Fe to the average of
the $\alpha-$elements vs [$\alpha$/H].
Upper panel: data for the Thin Disk stars;
middle panel: data for Dissipative Component stars;
lower panel: data for Accretion Component stars.
The mean line through various metallicity bins for Dissipative
component is overimposed in each plot.
The typical error bar is shown in the lower left corner}
\label{f:fig01}
\end{figure}

\begin{figure} 
\includegraphics[width=8.8cm]{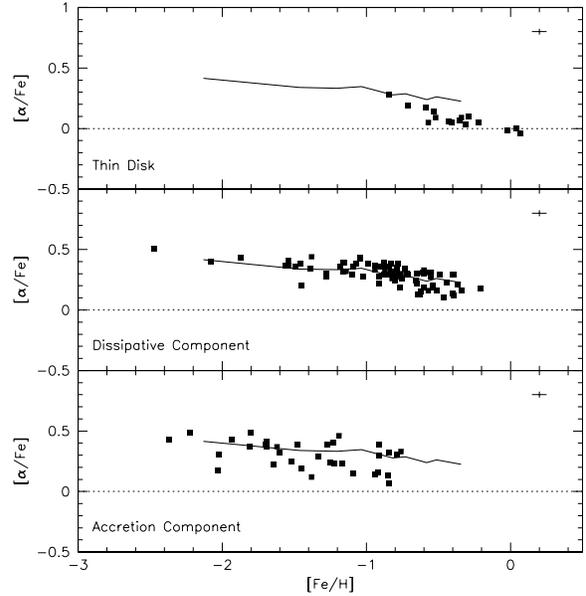}
\caption[]{ Run of the abundance ratio of the average of
the $\alpha-$elements to Fe vs [Fe/H].
Upper panel: data for the Thin Disk stars;
middle panel: data for Dissipative Component stars;
lower panel: data for the Accretion Component stars.
The mean line through various metallicity bins for Dissipative
component is overimposed in each plot.
The typical error bar is shown in the lower left corner}
\label{f:fig01b}
\end{figure}

\section{The Fe/$\alpha$\ ratio, and kinematics }

In Paper I we considered abundances of several $\alpha$-elements. In the
present discussion we will only consider four of these elements: Mg, Si, Ca,
and Ti. The most abundant among $\alpha-$elements, O, is not included
because its analysis is more difficult for a variety of reasons (only the IR
triplet at 7771-74~\AA\ could be used for the program stars; derived
abundances are very sensitive to errors in temperatures; they are affected by
significant departures from LTE; and finally they are quite sensitive to the
adopted model atmosphere, with significant changes in case plane-parallel
atmospheres are replaces by 3-d ones). We simply averaged the abundance ratios
to Fe given by the different individual elements (Ti abundances being
actually the average of abundances given by neutral and singly ionized
lines). Fig. 1 shows the correlation existing between
overabundances (with respect to Fe) of the various $\alpha-$elements, using
Ca as reference element: we selected Ca for this exercise because the
[Ca/Fe] is the abundance ratio most accurately determined from our data.
Different symbols are used for stars belonging to different populations.
Error bars for abundances of individual elements were determined in Paper I
(see Table 9 of that paper). For those elements that will be considered
throughout this paper, they are $\pm 0.05$\,dex for [Fe/H] and [$\alpha$/H];
$0.025$\,dex for the [Fe/$\alpha$] ratio (hereinafter $\alpha$\ represents the
average of Mg, Si, Ca, and Ti abundances); and 0.069\,dex for the [O/Fe],
0.055\,dex for the [Mg/Fe], 0.048\,dex for the [Si/Fe], 0.033\,dex for the
[Ca/Fe], and 0.030\,dex for the [Ti/Fe] abundance ratios.

A 45 degree degree line is a good match to the observed data in all panels of
Fig. 1, save for O. In this case, the slope is clearly
larger, as expected since virtually no O is expected to be produced by type
Ia SNe. The best fit line based on 66 stars (all but CD$-$80~328, that appears
to be discrepant) is 
${\rm [O/Fe]}=(1.72\pm 0.18){\rm [Ca/Fe]}+(0.089\pm 0.118)$, where the
error in the intercept represents the scatter of individual points. However,
as mentioned in Paper I, O abundances are more uncertain for binary or
reddened stars. Once these are eliminated, the best fit line is
${\rm [O/Fe]}=(1.80\pm 0.23){\rm [Ca/Fe]}+(0.046\pm 0.100)$. 
Offsets are present for the remaining elements. On average, the abundance
ratio to Ca are $0.132\pm
0.008$~dex (r.m.s. of 0.090~dex) for Mg, $0.002\pm 0.007$~dex (r.m.s. of
0.086~dex) for Si, and $-0.057\pm 0.006$~dex (r.m.s. of 0.074~dex) for Ti,
with no significant difference between stars belonging to the dissipative or
accretion components. Some of these systematic differences might be due to
offsets in the analysis; however, the systematic decrease of the ratio with
atomic weight corresponds well to the expected relative yields for different
classes of SNe, in particular to a systematic increase of the
contribution by thermonuclear SNe with increasing atomic weight among
the $\alpha-$elements; or to a systematic increase of the yields for the
lighter elements by core-collapse SNe with increasing mass. While
rather small, the r.m.s. scatter around the mean relations are
larger than predicted from the errors listed in Table 9 of Paper I: note that
what is relevant here are the errors in the abundance ratios to Ca, that are
expected to be 0.078, 0.055, 0.051, and 0.043\,dex 
respectively for O, Mg, Si,
and Ti. The r.m.s. scatters remain higher even if a few (two to five)
outliers are eliminated: in this case they are of 0.100\,dex for [O/Ca],
0.085\,dex for [Mg/Ca], and 0.067\,dex for both [Si/Ca] and [Ca/Ti]. Taking
into account the errors in the [Ca/Fe] ratio, there is still room for some
real star-to-star scatter in these ratios, of about 0.063, 0.055, 0.026
and 0.051\,dex for the [O/Ca], [Mg/Ca], [Si/Ca] and [Ti/Ca] ratios. 
While it is
well possible that these residual scatter are due to (underestimated)
observational errors, we note that again, there is a trend with atomic number
that may be suggestive of a real nucleosynthetic origin. However, the
conclusions of this paper will not depend on these considerations.

\begin{table}
\caption{Average [Fe/$\alpha$] values for the dissipative component in
various metallicity bins}
\begin{tabular}{crccr}
\hline
$[\alpha$/H$]$ & No.   & $<[\alpha$/H$]>$ & $[$Fe/$\alpha]$ & rms \\
bin (dex)      & Stars &     (dex)        &        (dex)    &  (dex)   \\
\hline
~~~~~~ - $-$1.3 &   4 & $-$1.715 & $-0.415\pm 0.040$ & 0.079 \\
$-$1.3 - $-$1.0 &   8 & $-$1.125 & $-0.339\pm 0.024$ & 0.067 \\
$-$1.0 - $-$0.8 &   7 & $-$0.869 & $-0.333\pm 0.020$ & 0.053 \\
$-$0.8 - $-$0.6 &  12 & $-$0.688 & $-0.346\pm 0.019$ & 0.065 \\
$-$0.6 - $-$0.5 &  15 & $-$0.544 & $-0.278\pm 0.016$ & 0.066 \\
$-$0.5 - $-$0.4 &  16 & $-$0.443 & $-0.286\pm 0.020$ & 0.078 \\
$-$0.4 - $-$0.3 &   8 & $-$0.342 & $-0.238\pm 0.030$ & 0.086 \\
$-$0.3 - $-$0.2 &  10 & $-$0.253 & $-0.261\pm 0.023$ & 0.074 \\
$-$0.2 - ~~~~~~ &   5 & $-$0.116 & $-0.226\pm 0.027$ & 0.061 \\
\hline
\end{tabular}
\label{t:tab01}
\end{table}

In the remaining part of this paper we will neglect the differences existing
between individual $\alpha-$elements (hereinafter not considering O),
and consider their average abundance. Fig. 2 displays the run of
the abundance ratio between Fe and the average of the $\alpha-$elements with
the average abundance of the $\alpha-$elements. Stars belonging to the
different components are shown in the three panels of this figure.
Fig. 3 displays similar plots, this time for the abundance
ratio between the average of the $\alpha-$elements and Fe with  respect to the
Fe abundance.

Table~\ref{t:tab01} lists the average abundances of the [Fe/$\alpha$] ratio
in different metallicity bins for the dissipative component. There is a clear
trend for increasing [Fe/$\alpha$] ratios with increasing metallicity:
the linear correlation coefficient is $r^2=0.257$\ over 85 stars, which
is significant at a very high confidence level. The following best fit
line seems to reproduce data very accurately:
\begin{equation}
[{\rm Fe}/\alpha] = (0.113\pm 0.021) [\alpha/{\rm H}] - (0.229\pm 0.071),
\end{equation}
where the error on the constant term represents the r.m.s. scatter along
the line. This scatter is about three times the internal error (0.026\,dex).

Part of this scatter is probably still due to analysis errors; however, we may
use it to set lower limits to the mass of clouds undergoing independent
chemical evolution during the formation of this component of the Milky Way.
Following the same approach described in Carretta et al. (2002), we will
assume for this exercise that the scatter around an average [Fe/$\alpha$]
ratio is due to the stochastic effect of a sampling of the initial mass
function by a discrete number of SN progenitors. Using the same precepts of
Carretta et al.,  we then obtain that typically about $\geq 100$~SNe 
contributed to the enrichment of each cloud. The exact value of this lower 
limit depends on the uncertain dependence of yields on the stellar initial 
mass. On turn, the value of the Fe abundance may be used to constrain the 
dilution factor of the ejecta of the SNe, and from the number of SNe and
the yields we may then get an estimate for the mass of the whole cloud, that
is of $\geq 10^5~M_\odot$. This value is not much different from what Carretta
et al. found for much more metal-poor stars, where the r.m.s. scatter but also
the dilution factor are larger. As noticed by the referee of Carretta et al.
(Bruce Carney),
this value for the mass of the clouds is of the same order of magnitude of the
Jeans mass at that epoch, and also of the same order of the mass of globular
clusters. This may be perhaps considered then as a typical mass for clouds
having an independent chemical evolution over a broad range of initial
conditions in our Galaxy.

We overplotted the line representing the average values of the [Fe/$\alpha$]
ratios for the dissipative component  also on the two other panels of
Fig. 2. Thin disk stars (shown in the upper panel) have a much
larger average value of the [Fe/$\alpha$] ratio at a given metallicity; the
mean residual with respect to the best fit line of Eq. (2) is $0.176\pm
0.014$\,dex, with an r.m.s of only 0.059\,dex, only slightly larger then the
internal error. This large difference in the [Fe/$\alpha$] ratios is simply
confirmation of the results obtained by Gratton et al. (1996, 2000) and
Fuhrmann (1998) for O and Mg. On the other side, stars belonging to the
accretion component show a much larger scatter than that observed for stars of
the dissipative component (the r.m.s. is of 0.108\,dex over 41 star). An F-test
shows that the difference between the scatter around the mean line for the
dissipative collapse and the accretion components is significant at 99.9\%
level of confidence, while a Kolmogorov-Smirnov test indicate that the two
distributions are different at 95.0\% level of confidence. Also the average
[Fe/$\alpha$] ratios of the two populations are different, in the sense that
the [Fe/$\alpha$] values of the accretion component at a given metallicity are
larger by $0.043\pm 0.017$\,dex; this is significant at 97.6\% probability
level (Student's {\it t} test). This result confirm an earlier finding (from a
much more limited sample) by Nissen \& Schuster (1997).

Finally, before interpreting the observed slopes in the [Fe/$\alpha$]
ratio as real, we should warn the reader that we performed an LTE analysis
using 1-d plane parallel model atmospheres. 3-D and non-LTE effects are
expected to depend on [Fe/H], and may affect abundances of Fe and 
$\alpha-$elements in different ways. Hence, a closer non-LTE/3-D study should
be undertaken before the important conclusion about type Ia SNe affecting the
chemical evolution of the dissipative component can be considered definitive.

\begin{figure*} 
\includegraphics[width=13cm]{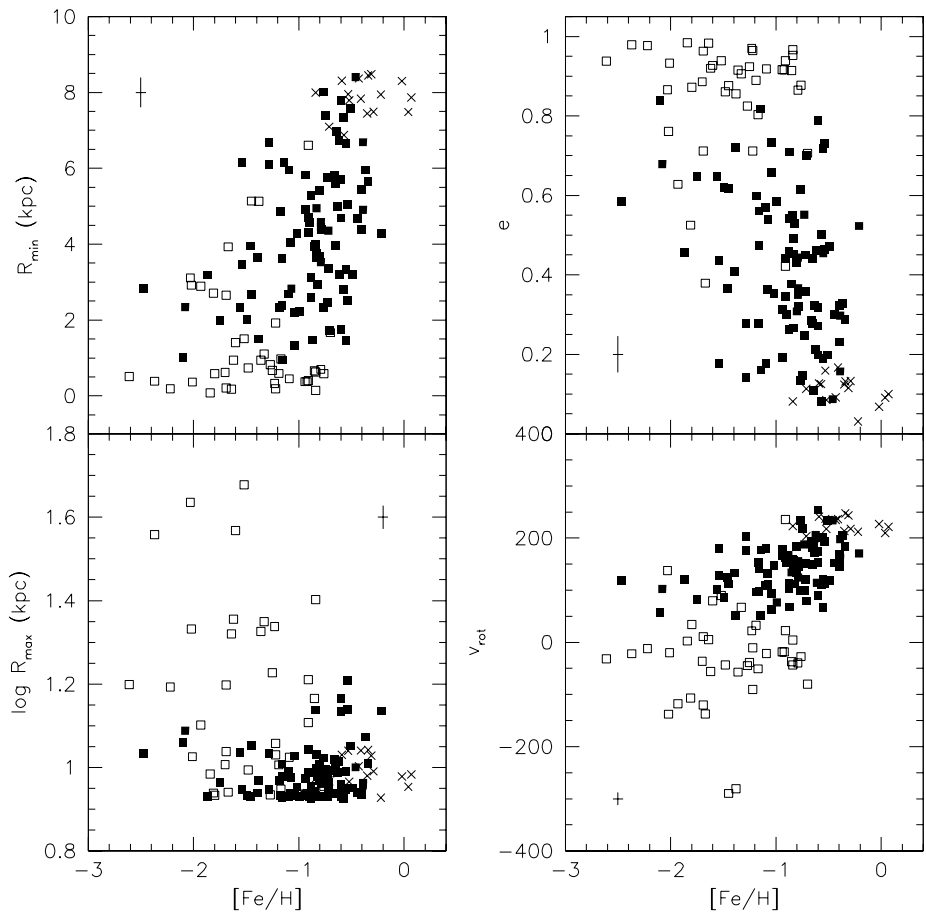}
\caption[]{ 
Run of the metallicity (from Fe) with various kinematical
parameters:
Perigalactic distance $R_{\rm min}$\ (upper left panel); Apogalactic distance
$R_{\rm max}$\ (lower left panel); Galactic orbit eccentricity $e$\ (upper
right panel); Rotational velocity around the Galactic center $v_{\rm rot}$\
(lower right panel).
Filled squares are data for Dissipative Component stars;
open squares those for the Accretion Component stars;
crosses are data for Thin Disk stars.
Typical error bars are shown in each panel}
\label{f:fig02}
\end{figure*}

\begin{figure*} 
\includegraphics[width=13cm]{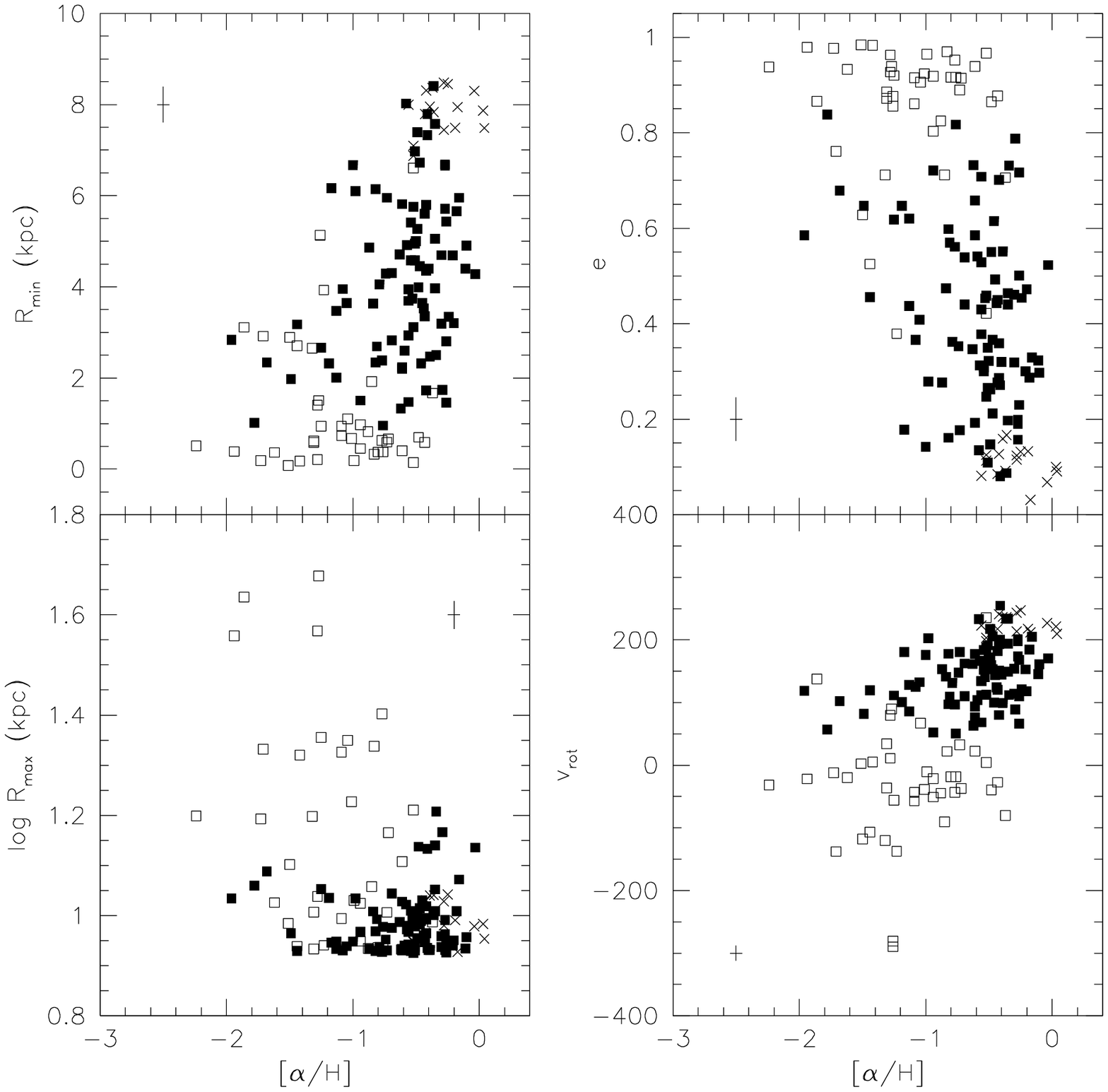}
\caption[]{ 
Run of the metallicity (from $\alpha-$elements) with various kinematical
parameters:
Perigalactic distance $R_{\rm min}$\ (upper left panel); Apogalactic distance
$R_{\rm max}$\ (lower left panel); Galactic orbit eccentricity $e$\ (upper
right panel); Rotational velocity around the Galactic center $v_{\rm rot}$\
(lower right panel).
Filled squares are data for Dissipative Component stars;
open squares those for the Accretion Component stars;
crosses are data for Thin Disk stars.
Typical error bars are shown in each panel}
\label{f:fig03}
\end{figure*}

\begin{figure*} 
\includegraphics[width=13cm]{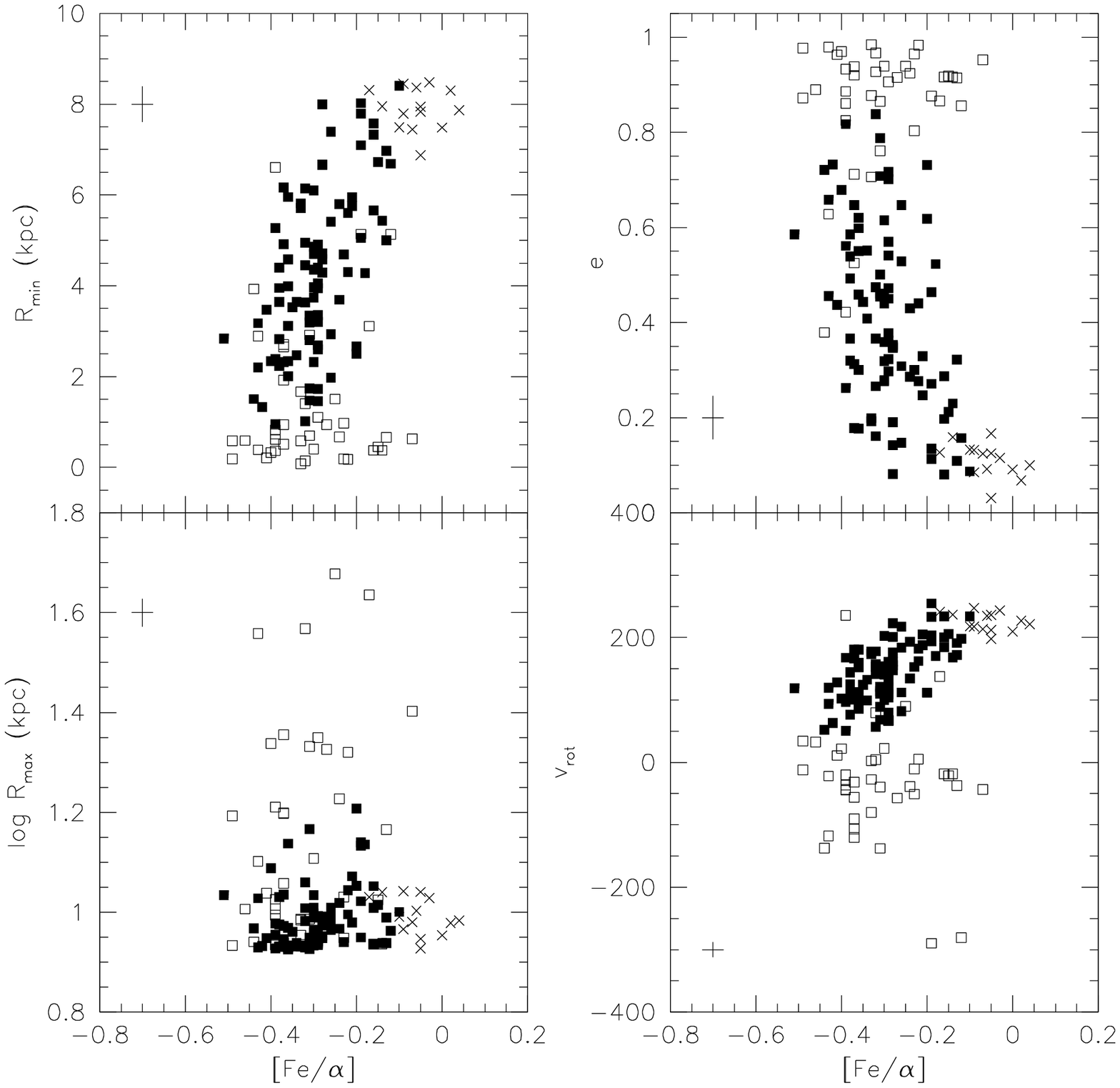}
\caption[]{
Run of the abundance ratio between Fe and $\alpha-$elements with various
kinematical parameters:
Perigalactic distance $R_{\rm min}$\ (upper left panel); Apogalactic distance
$R_{\rm max}$\ (lower left panel); Galactic orbit eccentricity $e$\ (upper
right panel); Rotational velocity around the Galactic center $v_{\rm rot}$\
(lower right panel).
Filled squares are data for Dissipative Component stars;
open squares those for the Accretion Component stars;
crosses are data for Thin Disk stars.
The typical error bars are shown in each panel as plus symbols}
\label{f:fig04}
\end{figure*}

\section{Correlation with kinematics and biases}

Correlating our abundances with {\bf kinematics} helps to understand several
important features of our sample of metal-poor stars. In Fig. 4
we displayed the run of the metallicity (from Fe) with various kinematical
parameters: Perigalactic distance $R_{\rm min}$\ (upper left panel);
Apogalactic distance $R_{\rm max}$\ (lower left panel); Galactic orbit
eccentricity $e$\ (upper right panel); Rotational velocity around the Galactic
center $v_{\rm rot}$\ (lower right panel). Different symbols are used for
stars belonging to the various populations considered throughout this paper.
Similar plots, but with (average) metallicity from the $\alpha-$elements are
shown in Fig. 5; while in Fig. 6 we plotted the
kinematical quantities against the abundance ratio between Fe and the
$\alpha-$elements.

Before opening a discussion on the results shown in these figures, we must
warn the readers about the biases that affect our results. There are mainly
three biases:
\begin{enumerate}
\item Our sample is drawn from the Hipparcos catalogue. This is complete only
down to approximately $V\sim 7$; fainter stars (the vast majority of those
included in the present sample) were obtained by combining lists proposed by
several groups. These included a rather large number of high proper motion
stars, so that there is a strong kinematical bias in our sample, favoring
objects on highly eccentric orbits, with low rotational velocities around the
galactic center, and with either large apogalactic or small perigalactic
distances;
\item Among the stars in the Hipparcos catalogue, we essentially selected
metal-poor stars, that is stars known to have [Fe/H]$<-0.8$\ (although a few
stars more metal rich than this limit are actually present in the sample).
The analysis of Carretta et al. (2000) shows that our sample is nearly
complete insofar stars with [Fe/H]$<-1.5$\ and accurate parallaxes are
considered, but it is far from being complete for more metal-rich stars.
However there is a clear bias against metal-rich stars;
\item Finally our sample only includes stars with accurate parallaxes; this
limits the volume sampled to a small region around the Sun, with a radius of
the order of 100\,pc, the actual value depending also on the absolute magnitude
of the stars. This biases our results against stars in the bulge, and also
against objects of the outer halo on orbits of low eccentricities.
\end{enumerate}
It is very difficult to properly quantify these biases. However their impact
on the results should be carefully considered in the following discussion.
Our results cannot be used to draw correlations between metallicity and orbit
parameters, a basic data that is much properly presented elsewhere (see e.g.
Sandage \& Fouts 1987, Norris 1994). Here, we will concentrate on a comparison
between distribution of abundance ratios among different populations, under the
assumption that they are less affected by the above mentioned selection
effects.

We note that in the case of the Dissipational Collapse Component the
rotational velocity is clearly correlated with the overall metallicity: the
linear correlation coefficients are $r^2=0.101$\ and 0.161 if [$\alpha$/H] and
[Fe/H] are used respectively (over 85 stars, that is 83 degrees of freedom).
Both these values are significant at more than 99\% level of confidence.
However, as mentioned above these correlations are at least in part artifacts
of the selection biases present in our sample. We think that the (closer)
correlation existing between the rotational velocity and the [Fe/$\alpha$]
ratio is much more interesting. In this case, the linear correlation
coefficient is $r^2=0.369$, over the same number of stars. This correlation is
significant at a very high level of confidence; as a matter of fact,
rotational velocity can be used to predict the value of the [Fe/$\alpha$]
ratio with an accuracy better than that obtained using overall metallicity:
the relation is:
\begin{equation}
[{\rm Fe}/\alpha] = (0.00109\pm 0.00016) V_{\rm rot} - (0.454\pm 0.065).
\end{equation}
The fact that the [Fe/$\alpha$] ratio is better correlated with the rotational
velocity than the overall metallicity is very interesting, because we expect
that the effects of selection biases are in this case only indirect; that is,
they should affect this correlation only through their impact on metallicity
and rotational velocity.

\section{Discussion and conclusion}

Our large sample of metal-poor stars with accurate abundance analysis and
kinematics allows a discussion of some of the chemical properties of the
components related to the dissipative collapse (rotating component) and to
accretion (either counter-rotating or not-rotating). When examining the plots
of Fig. 4-6, we should remind that $\alpha-$elements are mainly produced by
core collapse SNe. They then follow closely the star formation
history of the Galaxy. On the other side, since a significant fraction of Fe
is produced on much longer time scales by thermonuclear SNe, its
production is somewhat delayed with respect to that of the $\alpha-$elements.
Hence we expect that a Fe/$\alpha$\ ratio increasing with the time is a
signature of a lower star formation rate.

The first feature we notice in our results is that even if both 
the dissipative collapse and accretion components show a large overlap in
metallicity,  they show distinct chemical properties. In fact, while in both 
the cases there is a
significant deficiency of Fe with respect to the $\alpha-$elements, at a
given metallicity the Fe deficiency is more pronounced, and the scatter is
smaller, among stars belonging to the dissipational collapse component, in
agreement with earlier results by other authors (Nissen \& Schuster 1997). 
As far as the dissipational component is concerned, the Fe deficiency 
outlined above is a signature of a star formation rate higher than that of the
accretion component. Moreover the smaller scatter noticed above (Fig.4-6)
indicates that they formed from a more homogeneous interstellar medium that
the accretion component, may be well mixed over regions typically as massive
as $10^5~M_\odot$. Finally for the dissipational component we derive a close
correlation between chemical composition and kinematics, as expected for a
continuous metal enrichment during the dissipative collapse phase of the
Galaxy. However, while star formation occurred quite vigorously, it still left
time enough for a raise of the [Fe/$\alpha$] ratio from a value of
[Fe/$\alpha$]$\sim -0.40$\ at [$\alpha$/H]$\sim -1.5$, to a value of
[Fe/$\alpha$]$\sim -0.23$\ at [$\alpha$/H]$\sim 0$, where the evolution of
this component interrupts. This trend is very clear, but needs to be
confirmed by a 3-D/non-LTE study. As found by other authors (Gratton et al.
1996, 2000; Fuhrmann 1998; Reddy et al. 2003), this component is distinct from
the thin disk,  more rich in Fe, although shows an overlap in [$\alpha$/H].
The  raise of the [Fe/$\alpha$] ratio during formation of the dissipative
collapse component suggests that this lasted long enough to allow about 1/3 of
the Fe produced by thermonuclear SNe to be deposited in the interstellar
medium (again, with the caveat that this result might be due to
3-D/non-LTE effects not considered in the present analysis). While the time
scale for the evolution of the progenitors of these SNe is still not well
defined (see the review in Nomoto et al. 1999), this result suggests that the
collapse phase was perhaps not very short.

On the other side, the accretion component exhibits a much larger scatter in
the [Fe/$\alpha$] ratio: this is essentially due to the presence of a
significant population of Fe-rich stars, that made up nearly a third of the
sample. We prefer the notation Fe-rich stars rather than the usual one, 
$\alpha-$poor, because it is more suggestive of the origin of the anomaly:
these stars are in fact quite metal-rich, so that an interpretation of their
anomalies as due to random fluctuations of primordial SNe is not very likely
(it would require that these stars formed in small clouds with a typical  mass
of $3\,10^4~M_\odot$). Moreover within this framework we do not expect a
significant difference between the average [Fe/$\alpha$] ratio of the
accretion and dissipative collapse populations, at variance with what we find
here. It seems more plausible that these stars formed in an environment
with lower star formation rate and a chemical
evolution quite independent of that of the Milky Way, similar to that
found in the dwarf spheroidal galaxies, whose chemical composition has been
recently studied by Shetrone et al. (2003) and Tolstoy et al. (2003), that
halos tend to have higher Fe/$\alpha$\ than stars in the dissipative
component. Thus 
these stars should exhibit a large scatter in their ages.

As discussed in Sect. 4,
selection biases can affect  previous suggestions. Nevertheless
we wish to call the attention of
the reader to the last panel in Fig. 6, which shows the trend
of the [Fe/$\alpha$] ratio with rotational velocity. As discussed above, the
[Fe/$\alpha$] ratio is most likely an alias of time. While there is a clear
bias in this diagram in the sense that stars with low rotational velocity are
under-represented, there is no obvious bias at a given rotational velocity.
Thus it is interesting to compare the distribution of the values of the
[Fe/$\alpha$] ratio for stars belonging to the accretion component  with that 
for stars with moderate
rotational velocities ($40<V_{\rm rot}<150$\,\kms). While in the first group
roughly half of the stars have [Fe/$\alpha$]$>-0.3$\ (and so are probably
significantly younger), stars as rich in Fe are very rare in the second
group. The lack of  slowly co-rotating stars with high Fe/$\alpha$\ abundance  
ratios in the solar neighborhood is expected from dynamical models of disk 
formation. However a significant population of slowly counter-rotating Fe-rich
stars points toward  a long lifetime of their protogalactic clouds, in 
particular longer than that of the dissipational component. Moreover
the possibility that accretion component stars belong to the bulge and
that their kinematics properties have been changed by violent or two-body 
relaxation processes, is not appealing in view of their low metal abundance ([Fe/H$<-1$).

Finally, we wish to point out that perigalactic distances correlate with the
Fe/$\alpha$\ abundance ratios much better than the apogalactic distances. At
first sight, this result is quite surprising, since we would expect that the
apogalactic distance is a better signature of the region of formation of the
star and  that accretion component stars should form mainly in the
outer halo of the Galaxy. However, it should be noticed that the sample here
considered is drawn from the solar neighborhood, a region clearly distinct
from the outer halo. In order to observe in the solar neighborhood
stars formed in the outer halo,  highly
eccentric orbits are requested; if such an orbit passes very close to the 
galactic bulge, it
becomes chaotic, loosing most of the information about its initial conditions
(i.e. apogalactic distances).

\begin{acknowledgements}
This research has made use of the SIMBAD data base, operated at CDS,
Strasbourg, France; it was funded by COFIN 2001028897 by Ministero Universit\`a
e Ricerca Scientifica, Italy. We thank the referee for his helpful comments.
\end{acknowledgements}

\end{document}